\documentclass[a4paper,fleqn,preprint]{cas-dc_arx}

\usepackage[authoryear]{natbib}

\usepackage{tabularx}
\usepackage{multirow}

\def\tsc#1{\csdef{#1}{\textsc{\lowercase{#1}}\xspace}}
\tsc{WGM}
\tsc{QE}
\tsc{EP}
\tsc{PMS}
\tsc{BEC}
\tsc{DE}

\begin{document}

\let\WriteBookmarks\relax
\def\floatpagepagefraction{1}
\def\textpagefraction{.001}

% Short title
\shorttitle{Evaluating the impact ...}

% Short author
\shortauthors{J Podgorski et~al.}
%=================================================================                    

% Main title of the paper
\title[mode = title]{Evaluating the impact of the Central Chile Mega Drought on debris cover, broadband albedo, and surface drainage system of a Dry Andes glacier}
%=================================================================

\author[1]{Julian Podgórski}[orcid=0000-0001-5604-3462]

% Corresponding author indication
\cormark[1]

% Footnote of the first author
%\fnmark[1]

% Email id of the first author
\ead{jpodgo@igf.edu.pl}

\affiliation[1]{organization={Institute of Geophysics, Polish Academy of Sciences},
            addressline={ul. Księcia Janusza 64}, 
            city={Warsaw},
            postcode={01-452}, 
            country={Poland}}

% Second author
\author[2,3]{Michał Pętlicki}[orcid=0000-0003-3383-2586]
\ead{mpetlicki@udec.cl}
\affiliation[2]{organization={Jagiellonian University, Faculty of Geography and Geology},   city={Cracow},
            postcode={30–387}, 
            country={Poland}}
\cormark[1]

\affiliation[3]{organization={Universidad de Concepción, Department of Geography},
            city={Concepcion},
            country={Chile}}

% Third author
\author[3]{Alfonso Fernández}[orcid=0000-0001-6825-0426]

% Fourth author
\author[4]{Roberto Urrutia}[orcid=0000-0002-6239-9016]

\affiliation[4]{organization={Universidad de Concepción, Centro de Ciencias Ambientales EULA},
    city={Concepción},
    country={Chile}}

\author[5]{Christophe Kinnard}[orcid=0000-0002-4553-5258]
\affiliation[5]{organization={Université du Québec à Trois-Rivières},
    city={Trois-Rivières},
    state={Québec},
    country={Canada}}

% Corresponding author text
%\cortext[1]{Principal corresponding author}
\cortext[1]{Corresponding author}

\nonumnote{\\ \hspace{-0.8cm}
	\begin{tabularx}{0.5\textwidth}{lX}
		\raisebox{-.8\height}{\includegraphics[height=1cm]{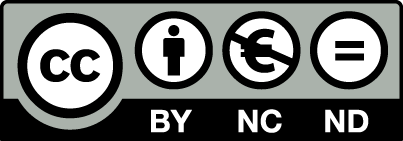}}
	 & \copyright~2023. This manuscript version is made available under the {CC-BY-NC-ND~4.0} license \url{https://creativecommons.org/licenses/by-nc-nd/4.0/} 
	\end{tabularx}
}

\begin{abstract}
In recent years, Chile has experienced an extraordinary drought that has had significant impacts on both the livelihoods of people and the environment, including the Andean glaciers. This study focuses on analyzing the surface processes of Universidad Glacier, a benchmark glacier for the Dry Andes.  Multiple remote sensing datasets are used alongside a novel spectral index designed for mapping of rock material located on the glacier's surface. Our findings highlight the precarious state of the glacier, which serves as a crucial water source for the region. The glacier exhibits locally varied debris accumulation and margin retreat. The most significant impacts are observed on the tongue and secondary accumulation cirques, with the latter at risk of disappearing. Debris cover on the tongue expands, reaching higher elevations, and is accompanied by glacier retreat, especially at higher altitudes. The equilibrium line is rapidly shifting upglacier, although mid-season snow cover tends to reach the 2013 equilibrium line even by 2020. Changes in stream density on the glacier's tongue indicate an increased water supply in this area, likely due to enhanced melting of glacial ice. These observed processes align well with meteorological data obtained from models. The behavior of dust and debris is influenced by precipitation amount, while the rate of retreat is linked to air temperature. 

%\noindent Each keyword shall be separated by a \verb+\sep+ command.
\end{abstract}

% Use if graphical abstract is present
\begin{graphicalabstract}
 \includegraphics{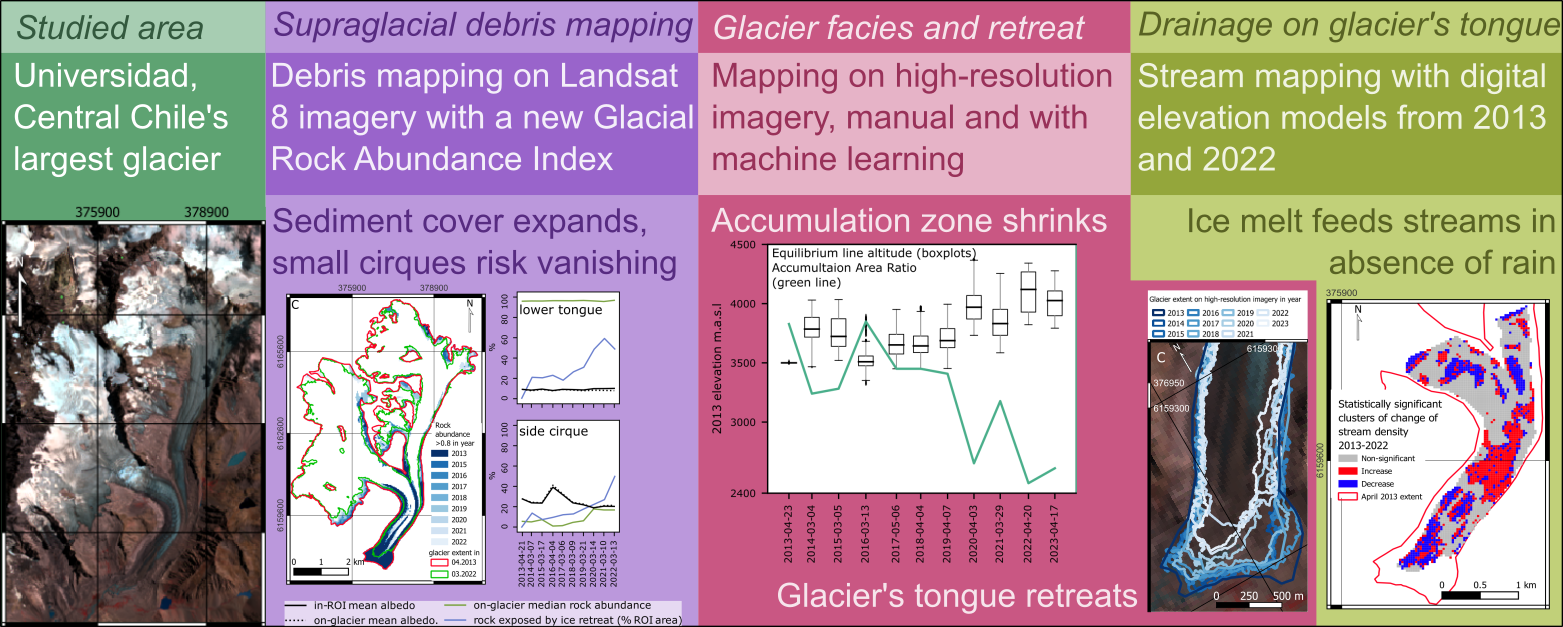}
\end{graphicalabstract}

 %% Research highlights
\begin{highlights}

\item The Glacial Rock Abundance Index (GRAI), a novel spectral index derived from Landsat 8 OLI images, can be effectively employed for mapping debris on the surface of a glacier.
\item Due to reduced precipitation the Equilibrium Line Altitude (ELA) is rapidly ascending, resulting in a reduction in the extent of exposed firn and leading to the gradual disappearance of the glacier's accumulation zone.
\item Universidad Glacier experiences an increase in the extent of surface debris cover in response to climate desertification.
\item The reduction of snow cover extent and presence of dust layer lead to lowering of the glacier's albedo
\item Localised transformations are found in the supraglacial drainage system in the ablation zone during the transition to a drier and warmer climate
\end{highlights}

% Keywords
% Each keyword is seperated by \sep

\maketitle

%%%%%%%%%%%%%%%%%%%%%%%%%%%%%%%%%%%%%%%%%%
%%
%% The article starts here
%%                                      
%%%%%%%%%%%%%%%%%%%%%%%%%%%%%%%%%%%%%%%%%%

\section{Introduction}

Contemporary climate change is profoundly influencing the global cryosphere~\citep{Vaughan2013}. Glaciers and ice caps are contributing strongly to sea level rise, despite their small mass relative to the ice sheets~\citep{Zemp2019}. These small ice bodies are predicted to contribute up to 20 cm of sea level rise by 2100, which accounts for approximately 1/6 of what is projected under the most pessimistic climate change scenario \citep{Slangen2016}. Additionally, it is estimated that up to 60\% of ice mass loss will occur among the glaciers within UNESCO World Heritage sites~\citep{Bosson2019}. Small glaciers not only suffer from the impacts of climate change but also serve as important indicators of its effects. This is evident from the inclusion of snow cover and glaciers on the list of key climatic variables by the Food and Agriculture Organization of the United Nations \citep[FAO,][]{Reuben2009}.

Research on small glaciers also holds significant societal relevance since in many regions worldwide, meltwater from mountain glaciers serves as a primary freshwater source and contributes to hydropower generation~\citep{Mark2015}. The thinning of these ice masses poses challenges to people dependent on them \citep{Baraer2012,Milner2017,Masiokas2020}. In the Dry Andes, glacier meltwater becomes an important part of the water supply during years with limited snowfall~\citep{Masiokas2020}. The loss of ice mass in the region is said to have mitigated societal and economical consequences of the drought for Chile~\citep{Dussaillant2019,Ayala2020}. 
The 2010-2018 period had particularly low precipitation in central Chile, sometimes called the Central Chilean Mega Drought~\citep{Garreaud2019,Kim2022}. The dry conditions have adverse impacts on water supply~\citep{McCarthy2022}, forests~\citep{Venegas2022,Urrutia2023} as well as on the glaciers of this part of the Andes. In the central region of Chile, drastic recession and disintegration of glaciers was observed between 1955 and 2013~\citep{Malmros2016}. Measurements of the Guanaco glacier, located in north-central Chile, revealed a negative mass balance since 2002~\citep{Kinnard2020}. Echaurren Norte Glacier, in Central Chile, has been loosing mass since 1993~\citep{Escobar1995,Masiokas2016}. The glaciers on the Argentinian side of the Andes have experienced consistent mass loss since the middle of the 20th century with a drastic acceleration between 2012 and 2020, a development blamed on the mega drought~\citep{Falaschi2022}.

\newpage
Ice and snow melting can take place across the entire glacier area. Supraglacial melt, which occurs at the glacier surface, is mainly related to the intensity of incoming solar radiation, ice and snow cover properties, and air temperature \citep{Hock2005}. \citet{Irvine2016} distinguish three important hydrological environments on a glacier surface: 1) snowpack with variable density and water permeability; 2) firn, functioning as an aquifer capable of retaining water coming from surface snowmelt; and 3) bare ice, the least permeable among them. These facies serve as water storage with different time scales, ranging from snow cover that captures and releases water on a seasonal scale, to retention in glacial ice for centuries~\citep{Jones2019}. The properties of bare ice may lead to formation of surface relief forms under flowing water through the process of incision and erosion along banks carved into the glacial ice \citep{Benn2010}. On temperate glaciers, the rate of stream incision frequently matches or surpass the ablation rate, and thus stream networks are created anew every melt season~\citep{Irvine2016}. The formation of drainage networks commences in the ablation season, after the snow cover of the ablation zone disappears~\citep{Fountain1998}. The on-glacier river networks assume a dendritic structure and resemble terrestrial river systems, characterized by meandering flows and the formation of pools in areas obstructed by topographic features. A wet ice surface has a lower albedo compared to a dry one~\citep{Ryan2018}, while varying and rough topography intensifies local melt processes~\citep{Cathles2011}. Combination of these two factors leads to increasing melt in areas of high channel densities and thus a feedback is established between accelerating melt and the densification of stream network~\citep{Pitcher2019}.

\citet{Brun2019} listed debris cover as one of the important morphological factors controlling glacier melt in the Himalayas, albeit not as important as slope and mean elevation. A thin cover of debris may increase melt rates~\citep{Oerlemans2009,Basnett2013}, but a sufficiently thick cover of sediments protects the underlying ice from elements and reduces melt~\citep{Ostrem1959,Vincent2016,Jones2019}. Climatic factors such as warming and reduced precipitation were hypothesized to promote expansion of debris cover on glaciers~\citep{Xie2020}. During the mega drought period, a decrease in surface albedo was observed across Chilean glaciers, particularly pronounced in the upper regions of the glacier, and with interannual fluctuations~\citep{Shaw2020}. However, this study was limited by its spatial resolution, preventing a comprehensive differentiation between the decline in albedo attributed to dust accumulation, expanding debris coverage, and glacier retreat. In the Dry Andes, several types of glaciers can be found, ranging from almost clear ice bodies, through debris-covered glaciers to rock glaciers~\citep{Janke2015}. Importantly, this classification is not static, and it is becoming more  common to observe gradual transitions between these categories as a consequence of the increasing debris cover resulting from the effects of climate change~\citep{Robson2022,Monnier2015}, carrying significant implications for meltwater production.

Studying the evolution of glacier surfaces, such as debris cover and supraglacial hydrology, requires accessible observations of glacier surface conditions. Satellite remote sensing has been used to map the surface and hydrology of glaciers. For instance, imagery acquired by the low-resolution MODIS has been employed to map lakes on the Greenland Ice Sheet~\citep{Selmes2011}. The surface drainage system of a glacier has been mapped using a~medium resolution dataset of Landsat imagery (30~m/pixel) by \citet{Wyatt2015}, whereas \citet{Smith2015}~have employed high-resolution (2~m/pixel) imagery from the WorldView-2 mission to map drainage patterns on specific sections of the Greenland Ice Sheet. 

A study on sub-pixel glacier classification by \citet{Yousuf2020} demonstrated the potential of spectral signatures, which are reflectivity patterns characteristic of specific surface types, for mapping different ice facies. The detection of diverse surface types on satellite imagery often relies on spectral indices, which establish relationships between the reflectivity of surfaces at different wavelengths to generate maps of specific features. \citet{Wessels2002} have employed the ASTER sensor (15 m/pixel) and several band-ratio indices to map supraglacial lakes in the Mount Everest area. Furthermore, \citet{Li2012} proposed an algorithm for glacial lake delineation based on the Normalized Difference Water Index (NDWI) band-ratio tailored to detect water on images~\cite{Gao1996}. Band-ratios also provde to be a valuable tool for mapping debris-covered glaciers. \citet{Shokory2023} employed custom spectral indices and Landsat 8 OLI/TIRS imagery, utilizing the panchromatic, thermal infrared, and blue bands, to detect sediment on the glacier surface in the Afghan Hindu-Kush region. A ratio of Near Infrared (NIR) and Shortwave Infrared (SWIR) bands has been successful in identifying exposed ice \citep{Haireti2016}. In a mapping workflow for a group of Chinese glaciers, \citet{Haireti2016} incorporated a ratio of Thematic Mapper bands 4 and 5 which cover near infrared (NIR) and short-wave infrared (SWIR) to detect clean ice areas. Additionally, \citet{Fleischer2021} used a band-ratio index of the NIR and SWIR bands from Landsat to detect clean-ice sections of glaciers and observed an upward migration of debris coverage on glaciers in the Alps. 

Outside the field of glaciology, a modified normalized differential index of Landsat~8~OLI NIR and SWIR was used to map karst landscapes~\citep{Pei2018}. The Karst Bare Rock Index (KBRI) proposed for this purpose modified the typical normalized differential formula by including a square root in the denominator~\cite{Pei2018}. The adjustment aims to enhance contrast and enables the detection of the amount of rock within pixels. To date, KBRI has been used for its intended purpose of mapping desertification in areas underlain by carbonate rocks~\citep{Alevkayali2023,Pu2021}, and has also served as a reference method for the development of remote sensing products~\citep{Ruan2023,Sales2023}. 

With the advent of high-resolution satellite imagery, pixel-based classification has been gradually replaced by Object-Based Image Analysis (OBIA), a paradigm of work with raster data that relies on segments identified within the image~\citep{Blaschke2014}. Each segment can be characterized by statistics of its pixel populations, including means, minima, and maxima of reflectivities or image textures. This approach offers a more extensive set of properties for analysis in comparison to pixel-based image analysis, where only band reflectivities and values of spectral indices from each individual pixel can be utilized. This is particularly advantageous for machine learning (ML) algorithms, as they generally perform better with a larger number of independent variables, albeit careful variable selection is required to avoid overfitting~\citep{Chen2020}. In glaciology, OBIA has been applied to map debris-covered glaciers~\citep{Rastner2014,Robson2015}, glacial lakes~\citep{Mitkari2017}, and even iceberg detection~\citep{McNabb2016}. Furthermore, the object-based approach has also been integrated with ML for studies of lacustrine icebergs~\citep{Podgorski2020}, as well as  glacier mapping in the Arctic~\citep{Ali2023} and the Pamir region~\citep{Lu2020}. However, the number of applications of this method in ice studies still remains low in comparison to the more classic pixel-based image analysis.

In this study, we present a decade-long analysis of the evolution of debris cover and supraglacial hydrology on a large glacier in the Dry Andes region – Universidad Glacier. Our approach involves a novel use of high-resolution hyperspectral imagery to create a model for detecting the glacier's debris cover within Landsat 8~OLI pixels. We introduce the Glacial Rock Abundance Index (GRAI) to quantitatively map and track the expansion of sediment cover and the retreat of the glacier. We assess the spatial variability of glacier facies derived from the application of object-based image analysis and machine learning, as well as the temporal changes in equilibrium line altitude (ELA). Finally, we examine the evolution of the surface albedo in response to its potential drivers, i.e. increasing sediment deposition, reduced snow cover and glacier retreat, and how supraglacial stream networks have changed between 2013 and 2022 in response to climate and glacier surface changes.  

\section{Study Area} 

\begin{figure}[ht] 
\centering
\includegraphics[width=\columnwidth]{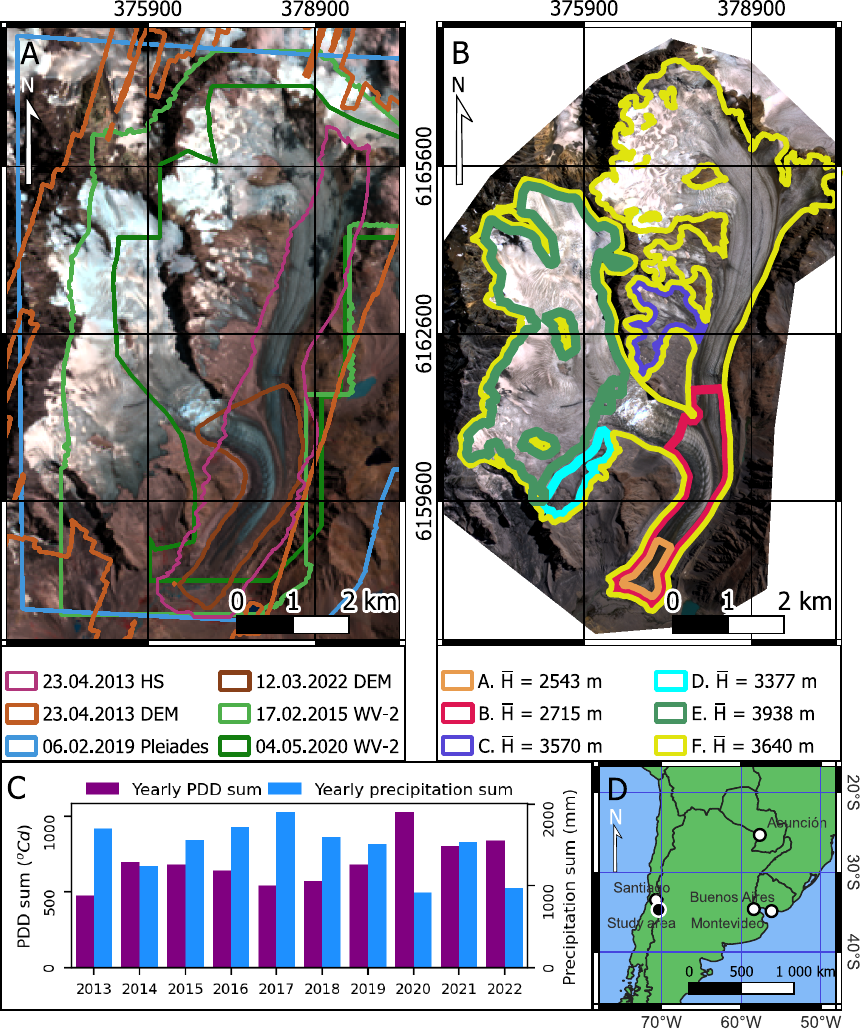}
\caption{Overview of the study area:
A -- Universidad Glacier with marked extents of the high-resolution datasets used. Background: 03.02.2023 Landsat 8 imagery;
B -- ROIs A-F with their mean 2013 elevation. Background: 30.03.2022 Dove imagery;
C -- Yearly PDD sum and precipitation sum for the study period sourced from the ERA5-Land model; 
D -- Study site location within the central part of South America.\label{fig:overviewMap}}

\end{figure}

Universidad Glacier is located in Chile, approximately 130~km south of the country's capital, Santiago, in the Dry Andes near the Chile-Argentina border (Fig.~\ref{fig:overviewMap}A). It is a valley glacier with a notable elevation range, spanning from 2450 to 4550~m~a.s.l. With a surface area of over 27~km\textsuperscript{2}~\citep{Kinnard2018}, it is also the largest glacier in Chile north of the Patagonian Icefields. 
Universidad Glacier is regarded as a benchmark glacier for the Dry Andes. The phenomena observed on such benchmark glaciers can be considered representative of an entire class of glaciers—in this case, mountain glaciers of the Dry Andes. Generally, larger glaciers tend to exhibit greater resilience than smaller ones to short-term climatic fluctuations~\citep{Paterson1994,Haeberli2005}. This makes Universidad Glacier a prime target for research of medium- and long-term processes. In contrast, the other glaciers in the region are smaller, which means that in their case, the low-frequency signals are at risk of being overshadowed by the high-frequency changes caused by year-to-year weather fluctuations.
The Chilean government acknowledged the glacier's status as a benchmark for assessing the impact of climate change, selecting it to represent its region in a country-wide survey of glaciers and designating it as a prime object of multidisciplinary studies~\citep{CECS2009}. Consequently, we regard Universidad Glacier as a representative example within the regional context. It provides valuable insights into the processes occurring on smaller glaciers of the Dry Andes region, which, due to logistical or financial constraints, have not undergone comprehensive surveying \citep{Bravo2017, Kinnard2018}.

\newpage
 In our work we analyze several regions of interests (ROIs) on Universidad Glacier, manually picked to represent diverse glacial environments (Fig.~\ref{fig:overviewMap}B). In the mid-20th century, \citet{Lliboutry1958} documented various features on the surface of Universidad Glacier, including glacier mills and ogives, and classified the glacier as Maritime (or "Temperate" according to modern classification). The presence of cold ice is not anticipated in the present times, given that the Chilean Andes have been undergoing a sustained warming trend since at least 1979, as noted by \citep{Falvey2009}. Consequently, the entire ice mass is expected to remain at the pressure melting point, thereby enhancing its susceptibility to melting.

Direct observations of glacier mass balance in the Dry Andes are limited \citep{Farias2019}. In the hydrological year 2012/13, the measured glaciological mass balance of Universidad Glacier was slightly negative, reaching  $-0.32\pm0.40$ m~w.e.~a$^{-1}$. However, in the following year~2013/14, the negative mass balance increased to -2.53$\pm$0.57~m~w.e~a$^{-1}$. This abrupt change was attributed to a combination of factors such as the amount of snowfall, surface albedo and cloud cover \citep{Kinnard2018}. A study based on satellite digital elevation models (DEMs) showed a long-term mass loss rate of 0.44$\pm$0.08~m~w.e.~a$^{-1}$ during the period of 2000-2013 \citep{Podgorski2019}. In the immediate vicinity of Universidad Glacier, there was a moderate mass gain observed between 2000 and 2009, followed by a period of intensive mass loss~\citep{Dussaillant2019}.

During droughts and dry summers, the glacier meltwater can be an important source of water for the densely populated central Chile~\citep{ValdesPineda2014}. One of the important rivers for Chilean water resources is Tinguiririca river, whose flow is partially sustained by Universidad Glacier located at the headwaters of the San Andr{\'e}s tributary~\citep{Kinnard2018}. An estimate provided by \citet{Bravo2017} indicated that as much as 20\% of flow in the river basin in 2010 was contributed by Universidad Glacier meltwater. Therefore, studies of the hydrology of this glacier can yield not only scientific advancements but also valuable insights for efficient local resource management.

%%%%%%%%%%%%%%%%%%%%%%%%%%%%%%%%%%%%%%%%%%
\section{Data}
In this study, we integrate various remote sensing products from different sources and types, including hyperspectral airborne imagery, very high- to medium-resolution multispectral satellite imagery, digital elevation models derived from airborne laser scanning and drone photogrammetry. These datasets are supplemented with meteorological and hydrological data to provide a comprehensive environmental context for interpreting the observed changes.
\subsection{Hyperspectral imagery} 
An airborne hyperspectral radiance image served as a basis to observe and quantify rock abundance on the glacier and relate it with the proposed new Glacier Rock Abundance Index (GRAI) applied to satellite optical imagery. The image was acquired on 23 April 2013 with a HySpex VNIR-1600 camera. The image covers the lower tongue of the glacier and the upper part of the main trunk of the glacier (Fig.~\ref{fig:overviewMap}C). The image has 160 spectral bands, of which 120 were used due to noise in the infrared bands (above 850~nm). 
\subsection{Multispectral optical satellite imagery}
For the detection of debris cover using the GRAI, we used reflectance images obtained from the Operational Land Imager (OLI) instrument mounted on the Landsat 8 satellite. A single image was manually selected per year spanning from 2013 to 2023. Our selection criteria prioritized scenes devoid of clouds and snow, captured during the late summer period, specifically in March or April. To handle the processing tasks, we harnessed the capabilities of the Google Earth Engine platform \citep{Gorelick2017} in conjunction with the Google Colaboratory online platform, employing the Python programming language for data sourcing, filtering, and manipulation.

The  glacier extent, accumulation areas and snow cover extent on the glacier were manually delineated from high-resolution satellite imagery originating from diverse remote sensing programs. WorldView-2 (WV-2) and Pleiades imagery were acquired as radiometrically and geometrically corrected, non-orthorectified datasets \citep[OrthoReady 2A processing level, ][]{DigitalGlobe2016}.  We applied atmospheric correction to both the WV-2 and Pleiades imagery, employing the ATCOR algorithm~\citep{Richter2017}, and subsequently pansharpened the images. RapidEye-1 images were acquired as the Top of Atmosphere (TOA) radiance, non-orthorectified images \citep[1B (basic analytic) processing level, ][]{DigitalGlobe2016}. The imagery from WV-2, RE-1 and Pleiades were orthorectified with the PCI Geomatica software, with the ASTER Global Digital Elevation Model v003~\citep{ASTERDEM2019} used as a source of a generalized reference model for terrain.
Additionally, we used imagery from Dove satellites, which are part of the PlanetScope constellation \citep{Planet2022}. The images were taken with the PS2, PS2.SD and PSB.SD instruments, and were acquired as orthorectified TOA radiance images (3B (analytic) processing level). Considering the diversity of instruments and the common purpose of these images, we will refer to them as Dove images hereafter.

\begin{table*}[htp]

\caption{Summary of imagery used in the study. BOA stands for Bottom of the Atmosphere, while TOA for Top of the Atmosphere. DEM stands for Digital Elevation Model.} \label{tab:Data}
\begin{tabularx}{\textwidth}{l|X|X|X|X}\hline

Sensor & Number of bands & Spatial resolution (m/pixel) & Processing level & Acquisition times \\ \hline

\multicolumn{5}{|c|}{Airborne datasets} \\ \hline
Riegl LMS-Q560 & 1 (elevation) & 1 m/pixel & DEM & 23.04.2013 \\ 
 EbeeX UAV DEM & 1 (elevation) & 5 cm/pixel resampled to 1 m/pixel & DEM & 12.03.2022 \\ 

HySpex VNIR-1600 & 160 & 1 m/pixel & at-sensor radiance & 23.04.2013 \\ \hline
\multicolumn{5}{|c|}{Satellite datasets} \\ \hline
PS2 & 4 & 1 m/pixel & TOA radiance & 2017, 2018 \\ 
PS2.SD & 4 & 1 m/pixel & TOA radiance & 2019-2021, yearly \\ 
PSB.SD & 8 & 1 m/pixel & TOA radiance & 2022, 2023 \\ 

RapidEye-1 & 4 & 3 m/pixel & TOA radiance & 2014-2016, yearly \\ 
Pleiades & 4 & 0.5 m/pixel & BOA reflectance & 06.02.2019\\ 
WorldView-2 & 8 & 0.5 m/pixel & BOA reflectance &  17.02.2015, 04.05.2020\\
Landsat 8 OLI & 7 & 30 m/pixel & BOA reflectance & 2013-2023, yearly \\\hline

\end{tabularx}
\end{table*}
\subsection{Digital elevation models}
We used two digital elevation models for mapping stream networks on the glacier tongue. The first one was acquired during the same measurement campaign as the hyperspectral dataset. The model is based on airborne laser scanning (ALS) with a Riegl LMS-Q560 laser scanner mounted on the airplane. The original resolution of this DEM is 1~m. It was described in more detail in a previous work on the elevation change of Universidad Glacier~\citep{Podgorski2019}. The second DEM is an UAV-imagery based product acquired on 12 March 2022. It was acquired with an EbeeX fixed-wing UAV equipped with an onboard double frequency GNSS receiver. The post-processing kinematic method was used to georeference the images to centimetric accuracy using a local GNSS base. The photogrammetric solution was calculated in Pix4d software, yielding a DEM with a 5~cm/pixel spatial resolution. This product was then resampled to 1~m/pixel to bring it to agreement with the earlier ALS DEM.

\subsection{Meteorological data} 
Hourly air temperature (T) and precipitation (P) above the glacier tongue were extracted from the ERA5-Land hourly reanalysis~\citep{Munoz2019} in GEE. Hourly air temperature and precipitation were extracted for the model cell that covers Universidad Glacier. 
ERA5 temperatures were extrapolated to each analyzed glacier ROI using a temperature lapse rate of -0.006\textsuperscript{o}C/m based on \citet{Bravo2017}. This gradient includes a constant value of -0.006\textsuperscript{o}C/m during nighttime and a generalized elevation lapse rate of -0.006\textsuperscript{o}C/m during the day, representing the mean value between the extremes of -0.007\textsuperscript{o}C/m and -0.005\textsuperscript{o}C/m mentioned by the authors. For each ROI, we used the temperature time series adjusted for elevation to calculate Positive Degree Days (PDD), which serves as a proxy for glacier ice melt. We calculated the annual PDD sum and annual cumulative PDD sum % maybe cumulative will go, we'll see
for the periods between April 1st and March 31st for each year from 2013 to 2022 (Fig.~\ref{fig:overviewMap}B). 
\subsection{Hydrological data}
 Mean monthly discharge values for the Tinguririca river, which receives meltwater from Universidad Glacier, were extracted thee database of the Chilean General Water Directorate. The data come from the gauge station located near La Virgen village, situated approximately 57~km downstream from the glacier terminus along the river course. As of 2023, there were two run-of-river hydropower generation installations located between the glacier and the gauging station. Notably, neither of these installations creates artificial lakes. Consequently, we assume that the discharge record obtained from the gauging station represents the natural flow of the river.

%%%%%%%%%%%%%%%%%%%%%%%%%%%%%%%%%%%%%%%%%%
\section{Methods}
\subsection{Mapping glacier surface facies}

We segmented the pre-processed Pleiades and WV-2 images using the multiresolution segmentation algorithm of eCognition 10.2~\citep{Trimble2017}. The segmented images were then classified with the Random Forest supervised classification algorithm. We manually selected polygons as training data based on visual interpretation of glacial features. Each image required a distinct training set due to significant differences in lighting conditions. Classification of all images included the following classes:  "ice" for exposed glacial ice; "blue ice" for segments of deep, non-metamorphised glacial ice exposed in crevasses; "dark ice" for exposed ice with embedded sediments, most often found on the glacier tongue; "snow/firn" for areas of snowpack and exposed firn, as the two are not distinguishable on the available imagery; "rock" for rocks surrounding the glacier and debris accumulated on the glacier surface; "deep shadow", a class intended to filter out the darkest shadows so that they do not pollute the classification of other classes; "glare" for very bright areas, saturated with solar glare, found only on the 2020 WV-2 image.
The properties used for object classification were the mean reflectances of each image band, as well as the means of several spectral indices, and four Haralick texture measures: Angular 2\textsuperscript{nd} momentum, Entropy, Contrast, and Standard Deviation~\citep{Haralick1973}.  

We used Dove, WV-2, RE-1 and Pleiades imagery to manually map the outlines of the glacier and edge of the snow cover in the subsequent years. We manually traced the edge of the glacier within the eCognition 10.2 program using segmented multispectral imagery. This process considered the demarcation between the debris-covered ice and the surrounding valley slopes or sediment-covered areas resulting from glacier recession. The outcome of this effort was a set of 11 polygons representing the glacier extent for the years 2013-2023. 

The equilibrium lines were digitized as the lower edge of the snow cover at a date close to the end of the ablation season, as visually identified in the image. Equilibrium Line Altitudes (ELA) were calculated as the mean of the 2013 DEM elevations sampled along the digitized line. We compared the ELA values with seasonal (3-month) Oceanic Ni{\~n}o Index (ONI) values for winter (June, July, August; JJA) and summer (December, January, February; DJF) sourced from NOAA~\citep{NOAA2023}. This value served as a proxy of the El~Ni{\~n}o Southern Oscillation (ENSO) stages. We calculated Spearman correlation between ELA, summer ONI and previous year's winter ONI, and determined the correlation's significance with a permutation test.

\subsection{Mapping sediment cover evolution}
A rock abundance map that represents the proportion of rock material within each pixel across the section of the glacier covered by the hyperspectral (HS) data was generated. This includes both the accumulation of sediment on the glacier surface and sections of rock exposed due to the glacier retreat. We created a spectral library by manually selecting group reference pixels (endmembers) for each of the two defined land cover classes (bare ice and bare rock) within the HS image. For the unmixing task, we used MESMA (Multiple Endmember Spectral Mixture Analysis) instead of simple linear unmixing, as MESMA can handle multiple endmembers per class~\citep{Roberts1998}. We employed an implementation of MESMA specifically designed for the Quantum GIS software program~\citep{Crabbe2020} because it incorporated a third category called "Shade," which accounted for the influence of varying lighting conditions on the appearance of areas covered by identical materials. 
The percentages of "HS ice" and "HS rock" classes were adjusted to ensure that their  sum equaled 100\%, while preserving the initial proportionality between "HS ice" and "HS rock" abundances. We used the averaging algorithm provided by the GDAL package (GDAL\_Average) to aggregate the HS rock abundance map to the resolution of the OLI imagery~(30~m/pixel). 

The Glacial Rock Abundance Index (GRAI) was used as a proxy to assess the presence of rock material on the glacier surface and its vicinity. The index is calculated using the Formula~\ref{eq:KBRI}: 

\begin{equation}\label{eq:KBRI}
    GRAI = \frac{\rho_{SWIR1}-\rho_{NIR}}{\sqrt{\rho_{SWIR1}+\rho_{NIR}}}
\end{equation}

where $\rho$ stands for surface reflectance of the indicated OLI band. We included the square root of the sum of the two bands to enhance the contrast within the resultant index image, thereby facilitating the quantitative mapping of sediment presence on the glacier's surface. The formula is similar to the KBRI proposed by \citet{Pei2018} for the mapping of exposed karst rocks. However, GRAI is normalized to yield values between -1 and 1, in contrast to KBRI, which was overly tuned to a specific site and rock type. To devise a novel method to map the glacier debris cover, we paired HS data with a map of GRAI index created with an OLI scene acquired at the time closest to the acquisition of the HS image. 
The selected OLI image was taken on 21 April 2013, two days before the HS image. Meteorological data obtained from the ERA5-Land model indicate that there was no precipitation recorded in the area of Universidad Glacier during the time period between the two image acquisitions. 
Subsequently, we developed a statistical model to convert OLI GRAI into surface rock abundance by fitting a generalized logistic curve between OLI GRAI and hyperspectral-based rock abundance. Using this model, we calculated the rock abundance for each OLI image in our time series. In addition to GRAI, we calculated broadband albedo using the same OLI imagery, using the formula established by \citet{Liang2001}:

\begin{equation} 
    \label{eq:albedo}
     \begin{aligned}
    \alpha_{OLI} = 0.356\rho_{blue}+0.130\rho_{red} + 0.373\rho_{NIR}\\ + 0.085\rho_{SWIR1} + 0.072\rho_{SWIR2}- 0.018
    \end{aligned}
\end{equation}

where $\rho$ stands for pixel's reflectivity in the indicated bands.

Within each of the manually selected ROI we calculated the following timeseries: percentage of non-glacierized surface within the ROI; rock abundance and albedo for the glacierized area within the ROI;  annual and cumulative PDD calculated using elevation-corrected ERA-5 Land temperatures.

To analyze the relationships between the various measures of surface darkening or glacier retreat and the meteorological variables obtained from the ERA5-Land model, we constructed Spearman correlation matrices for each of the ROIs. We chose to use the Spearman coefficient due to the significant non-normality of the rock abundance distributions within the ROIs. To determine the statistical significance of these correlations, we performed a permutation test using the scikit-learn Python library. 

For visualisation and examination of expansion of debris cover and glacier retreat we created vector maps of the debris cover in years 2013-2022. 
We applied the Accelerated Hierarchical Density-Based Clustering Algorithm \cite[HDBSCAN, ][]{Mcinnes2017} unsupervised machine learning algorithm, an extension of the DBSCAN algorithm \citep{Campello2013}, to each of the GRAI-derived sediment cover maps.  %to perform 3D clustering. 
We converted thus created clustered data to vector maps and used them to study the expansion of debris cover during the 2013-2022 period. The manually traced outline of Universidad Glacier on the high-resolution hyperspectral (HS, 2013) image was used for calculations on the whole glacier.

\subsection{Mapping changes of drainage network on the glacier's tongue}
We generated maps of streams in the lower section of the glacier tongue on the 2013 and 2022 DEMs. The extent of the stream map is constrained by the extent of the more limited 2022 DEM. In order to accurately include glacial wells in the drainage system mapping, we created a well map by identifying local depressions in the elevation models. We used the PCRaster plugin~\citep{Karssenberg2010} in QGIS 3.28 software to generate Local Drain Direction (LDD) maps for both DEMs with the wells map used as a representation of water sinks. Using this information, we derived maps of stream orders, which served as the basis for the general stream network map. In order to remove data artifacts, streams with a length of 1~m or less were removed from the final map.

We used square polygons aligned with the Landsat 8 pixel grid to derive stream density (m/m\textsuperscript{2}) and mean stream order. Afterwards, we calculated Moran's~I~Local Index of Spatial Association (LISA) on the difference map of stream order and density (2022 minus 2013) to search for local changes between 2013 and 2022. For Moran's~I~LISA we used the GeoDa software~\citep{Anselin2006,Anselin2018}. The resulting maps of the LISA value represents the statistical significance of clustering in change signals, compared to a random spatial distribution of values. Additionally, the data was classified into interpretative classes based on clusters of high/low difference values. 
%%%%%%%%%%%%%%%%%%%%%%%%%%%%%%%%%%%%%%%%%%
\section{Results} 

\subsection{Multispectral mapping of glacier facies}

\begin{figure*}[ht]
\includegraphics[width=\textwidth]{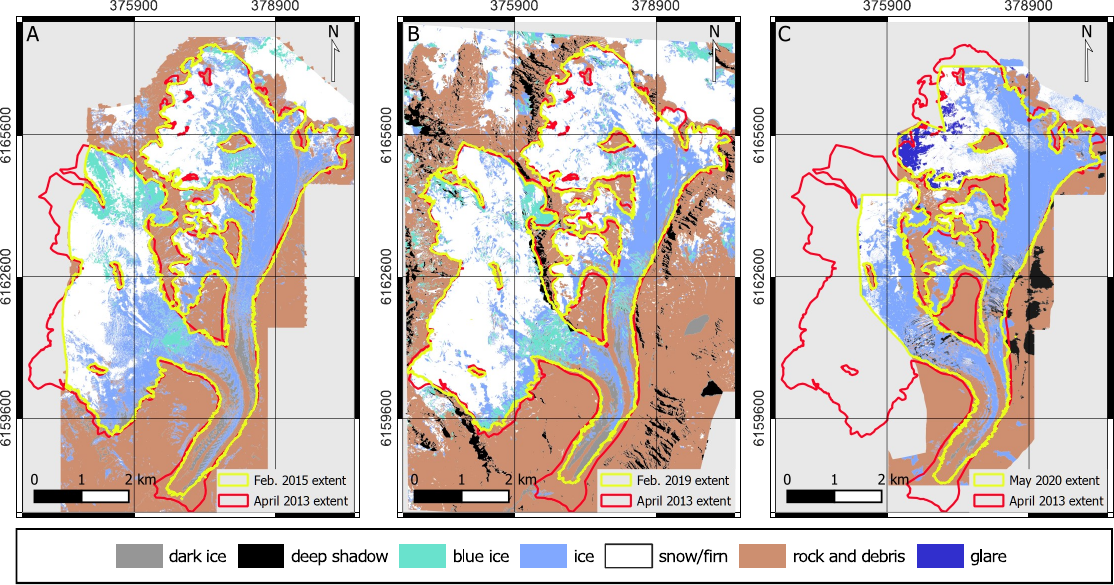}
\caption{Surface cover type on Universidad Glacier determined through  object-based classification of high-resolution multispectral satellite imagery. A -- WorldView-2 image, 17.02.2015; B -- Pleiades image, 06.02.2019; C -- WorldView-2 image, 04.05.2020. \label{fig:MSfacies}}
\end{figure*} 

\begin{table}[h]
\caption{Classification accuracies and kappa values of OBIA classification of high-resolution satellite imagery\label{tab:obiaAccu}}
\begin{tabularx}{\linewidth}[t]{lXXX}\hline

Dataset & 2015 WV-2 & 2019 Pleiades & 2020 WV-2 \\ \hline
Classification accuracy & 0.98 & 0.96 & 0.96 \\
Kappa value & 0.97 & 0.93 & 0.94 \\ 
\end{tabularx}
\end{table}

Object-based classification of high-resolution imagery was successful, returning high classification accuracy (>0.90) as measured with the samples used for training (Tab.~\ref{tab:obiaAccu}). 

The "snow" class represents areas covered by a smooth, very bright, white surface. This class includes both snow and firn, as they were visually and algorithmically indistinguishable from each other.  In the upper reaches of the glacier, snow or firn penitentes were observed. Despite being visually colored by a layer of sediment dust, the snow cover is easily distinguishable from the surrounding rocks and the accumulated supraglacial sediment cover. The "fresh ice" class, trained to detect blue-tinted ice exposed within icefalls, also appears on bright objects in the highest parts of the accumulation zones, emphasizing the spectral similarity between the bright species of snow and ice (Fig.~\ref{fig:MSfacies}). "Ice" class represents exposed surface ice of the glacier, not covered by the snow or firn. The class includes clean surface ice and its weathered species, as well as the surface of ice covered by a thin layer of dust. The "rock" class in turn covers the rock surface of the slopes surrounding the glacier and areas of thick moraines on top of the glacier in areas where sediment forms a thick layer without visible ice. 

In 2019 the mid-season snowline reached lower than in 2015, approaching the western icefalls and extensively covering the small side-cirques. In 2020, the snowline retreated higher, and areas of the glacier that were covered by snow in 2015 and 2019 revealed exposed ice.

The sequence of multispectral facies classifications shows a gradual darkening and expansion of sediment cover on the glacier tongue (Fig.~\ref{fig:MSfacies}). The use of high-resolution imagery enables distinguishing between the dark ice of the ogive bands and the actual accumulation of sediment.
A strip of still-exposed ice is evident in the eastern part of the tongue, realized as area of "dark ice" class. This class includes areas of exposed ice with significant amount of sediment embedded making the ice dark. It was necessary to include it as this species of exposed ice is distinct spectrally from both thick sediment cover and clear, exposed ice.

In areas of ice recession in the accumulation zone -- southwest margin of the western accumulation zone and the two small accumulation basins -- the newly exposed rock was correctly classified as rock material, rather than darkened ice. A gradual narrowing of the ice outlets from the small cirques to the glacier's main trunk is visible on the WV-2 and Pleiades imagery as expansion of the "rock" class, accounting for both retreat and accumulation of sediment in these narrow ice corridors (Fig.~\ref{fig:MSfacies}).

The 2020 WV-2 classification required slightly different approach due to presence of limited, semi-transparent cloud cover and significant brightness differences between areas exposed to direct sunlight and those covered by shadows. It was necessary to include a "glare" class which covered areas of saturated pixels in the uppermost reaches of the glacier (Fig.~\ref{fig:MSfacies}C). The clouds and deep shadows warranted use of sub-classes "shaded ice", "clouded snow" and "clouded ice" which were subsequently rolled into "ice" and "snow" classes.

\subsection{Manual mapping of ELA}

\begin{figure}[h]
\includegraphics[width=\columnwidth]{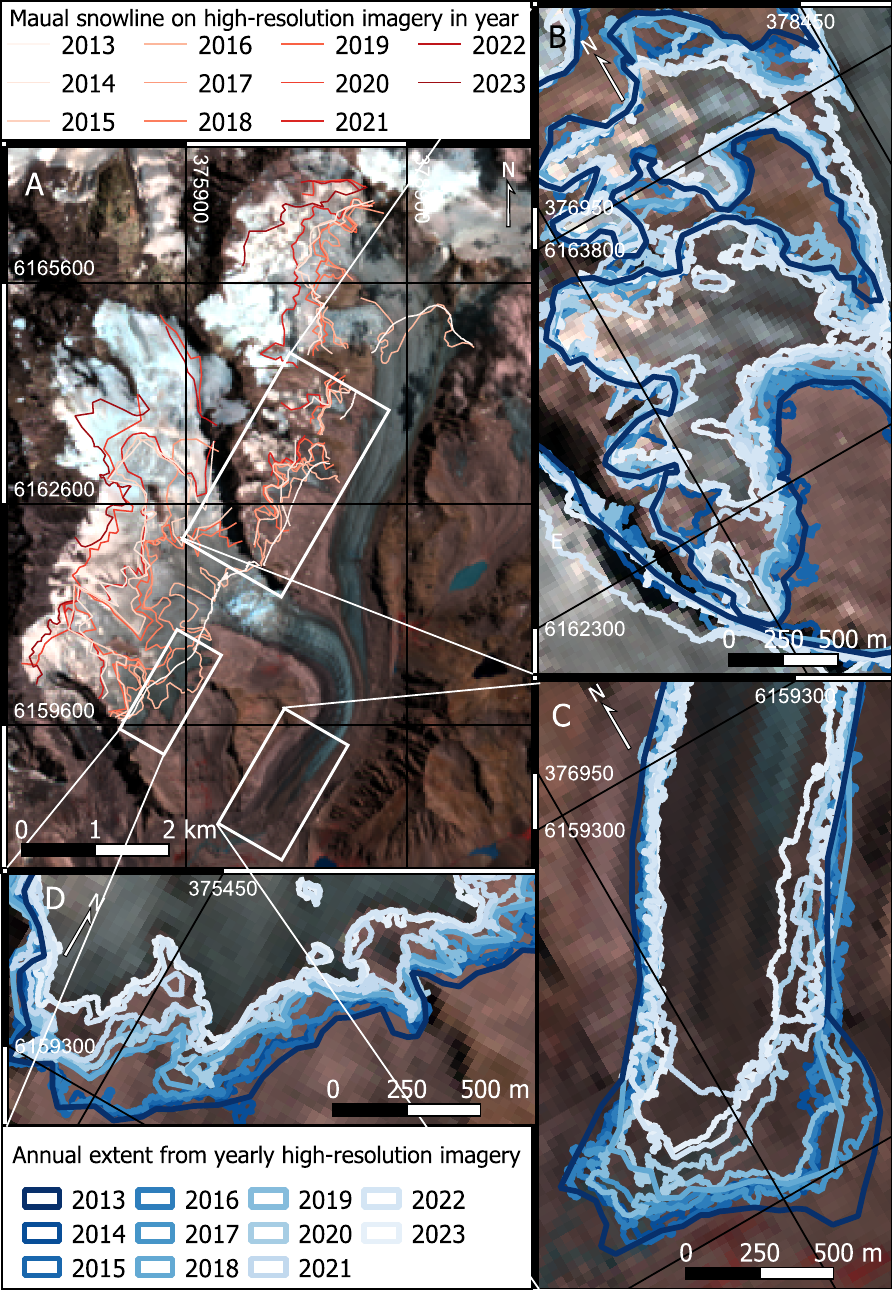}
\caption{Universidad Glacier extent manually digitized on high-resolution Dove satellite imagery in 2013-2023: A -- overview with snow extent; B -- recession of the small cirques; C --  recession of the glacier's tongue; D -- recession in the south-eastern part of the western accumulation zone.
} \label{fig:resultsMap}
\end{figure}

\begin{figure}[h]
\includegraphics[width=\columnwidth]{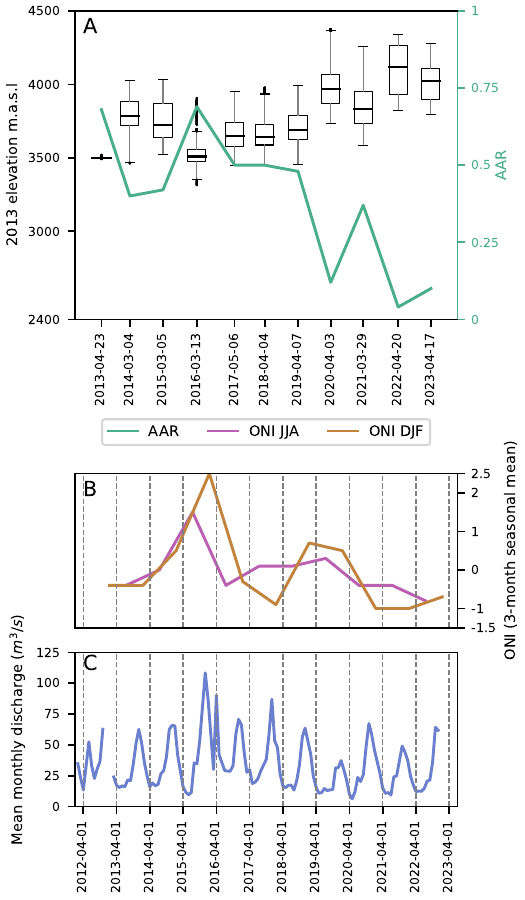}
\caption{ELA, AAR and ONI changes compared to hydrological record of Tinguririca river. A -- ELA values with 1~$\sigma$ error bars (black boxplots), and AAR values in each year (green line); B -- ONI seasonal values for winter (brown) and summer (violet); C -- Mean monthly discharge measured in Bajo los Briones gauge station. Note vertical lines on panels B and C, indicating the end of hydrological year on 1 April of each year.}\label{fig:ELAgraph}
\end{figure}

\begin{table}
\caption{Equilibrium Line Altitude (ELA) and Accumulation Area Ratio (AAR) of Universidad Glacier in 2013-2023. The values are based on manual delineation of the equilibrium line in optical satellite imagery for years 2014-2023, whereas for year 2013 the value reported by \citet{Kinnard2018} with no error bounds is provided. The two rightmost columns list 3-month ONI values for summer and previous year's winter.\label{tab:ELAAAR}}
\begin{tabularx}{\linewidth}[t]{lXXXX}

\hline
Year & ELA$\pm$1$\sigma$ (m.a.s.l.) & AAR & ONI Summer (DJF) & ONI prev.winter (JJA) \\ \hline
2013 & 3500         & 0.68 & -0.40 & 0.20 \\ 
2014 & 3797$\pm$121 & 0.40 & -0.40 & -0.40 \\ 
2015 & 3756$\pm$138 & 0.42 & 0.50  & 0.00 \\ 
2016 & 3526$\pm$93  & 0.69 & 2.50  & 1.50 \\ 
2017 & 3661$\pm$110 & 0.50 & -0.30 & -0.40 \\ 
2018 & 3669$\pm$108 & 0.50 & -0.90 & 0.10 \\ 
2019 & 3710$\pm$110 & 0.48 & 0.70  & 0.10 \\ 
2020 & 4009$\pm$173 & 0.12 & 0.50  & 0.30 \\ 
2021 & 3853$\pm$141 & 0.37 & -1.00 & -0.40 \\ 
2022 & 4105$\pm$169 & 0.04 & -1.00 & -0.40 \\ 
2023 & 4007$\pm$120 & 0.10 & -0.70 & -0.80 \\ \hline
\end{tabularx}
\end{table}

The ELA, as traced by the extent of snow/firn cover, has globally increased since 2012, wiht a marked acceleration  in recent years (Fig.~\ref{fig:resultsMap} and~\ref{fig:ELAgraph}A). The 2013 ELA of 3500~m.a.s.l. was surpassed by more than 200 meters the following year and remained above 3700~m.a.s.l. until 2023, except for the period from 2016 to 2018. During this period, the snow cover retreated up to 3526$\pm$93~m.a.s.l. in 2016 and above 3660~m.a.s.l. in 2017 and 2018 (Tab.~\ref{tab:ELAAAR}). In 2020, the ELA exceeded 4000~m.a.s.l. (4009$\pm$173~m.a.s.l.) for the first time, and the AAR dropped to 0.12.

The reduction in the AAR corresponds to the diminishing extent of snow/firn cover (Fig.~\ref{fig:ELAgraph}A). While the AAR was 0.68 in 2013, it dropped below 0.50 as early as 2014, paralleling the increasing ELA. By 2023, the AAR had reached 0.10, which is less than one-fifth of its initial value in 2013 (Table~\ref{tab:ELAAAR}). The year 2022 marked the highest ELA and the lowest AAR.  During this year, the lower edge of the snow extent reached an elevation of 4105$\pm$69~m.a.s.l. resulting in only 4\% of the glacier's area remaining in the accumulation zone.

Out of the four Spearman correlations computed between ELA, AAR, and ONI, only the relationship between the AAR and the ONI of the previous year winter was statistically significant (r = 0.60, $p$<0.05). The remaining correlations were lower and not significant, i.e. r = -0.43 for the ELA with the summer ONI and r = -0.48 with the previous year winter ONI. The AAR was correlated with the summer ONI at r=0.50.

\subsection{Evolution of rock abundance}

\begin{figure}[h]
\includegraphics[width=\linewidth]{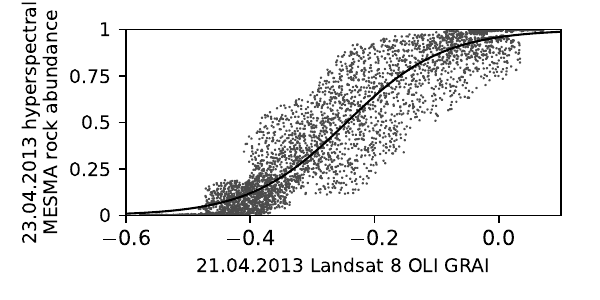}
\caption{Logistic regression model of relationship between the sediment coverage of hyperspectral pixels and the value of Landsat 8 OLI GRAI. \label{fig:logisticmodel}}
\end{figure}

The GRAI values are negative in all areas of the hyperspectral image extent, except those most heavily covered by debris or exposed rock (Fig.~\ref{fig:logisticmodel}). The relationship between hyperspectral rock abundance and OLI GRAI generally follows a linear pattern in the middle section but exhibits wide clusters of points near the extremes (0 and 100\%). Notable deviations from the linear pattern can be observed at both extreme ends, with distinct ranges of GRAI values extending beyond the main cluster of value pairs. The logistic curve provides a good fit to the dataset, capturing both the linear-like trend observed in the intermediate values of debris cover and the wide ranges of L8 GRAI values assigned to MESMA pixels that are either fully covered with rock material or not rocky at all~(Fig.~\ref{fig:logisticmodel}).

\begin{figure*}[ht]
\includegraphics[width=\textwidth]{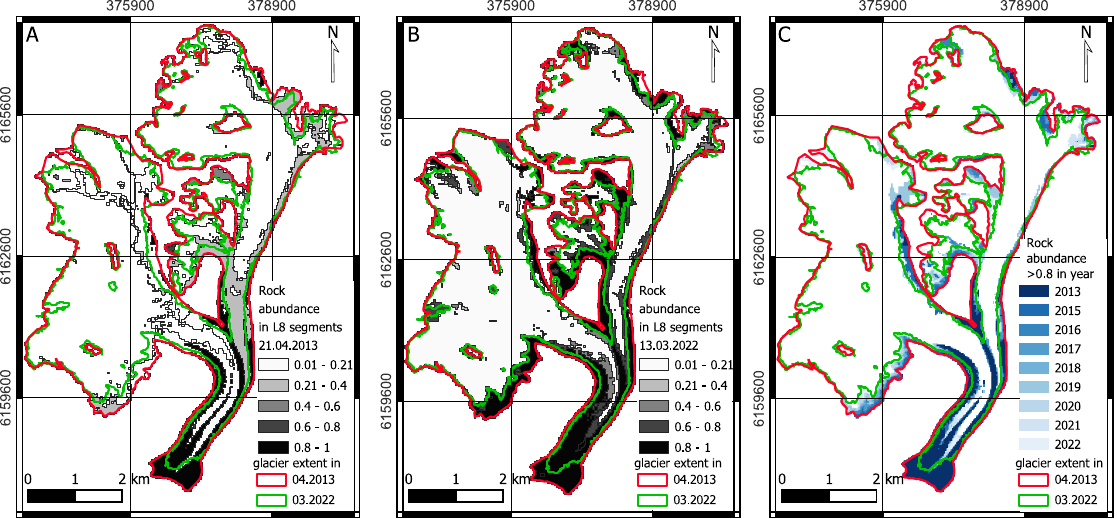}
\caption{Segmented Landsat 8 maps of rock abundance of Universidad Glacier. A -- rock abundance on 12.04.2013; B -- rock abundance on 13.03.2022; C -- segments with sediment coverage higher than that observed between the years 2013-2022 overlaid to illustrate the expansion of high-rock abundance areas over time. \label{fig:L8debris}}
\end{figure*}

%\subsection{Evolution of sediment cover}
 In 2013, the only noticeable features on the otherwise bright ice surface were the medial and lateral moraines in the lower part of the tongue, along with thin layers of sediment %  patches of darkness 
 in the western accumulation zone and above the eastern icefall (Fig.~\ref{fig:L8debris}A). In 2022, on the other hand, multiple zones of continuous sediment coverage can be observed on the glacier tongue, along the sides of the trunks, and at the edges of accumulation zones (Fig.~\ref{fig:L8debris}B). Most notably, the southern section of the western accumulation basin and the smaller cirques to the west of the glacier main trunk show increasing rock abundance.  A visible increase in mapped rock abundance is seen over time in the glacier's tongue and the select zones of accumulation zone (Fig.~\ref{fig:L8debris}C). The expansion can be attributed to ice retreat along the glacier margins and to expanding debris cover within the glacier.

\subsection{Retreat and sediment accumulation within ROIs}

\begin{figure}[h]
\includegraphics[width=\linewidth]{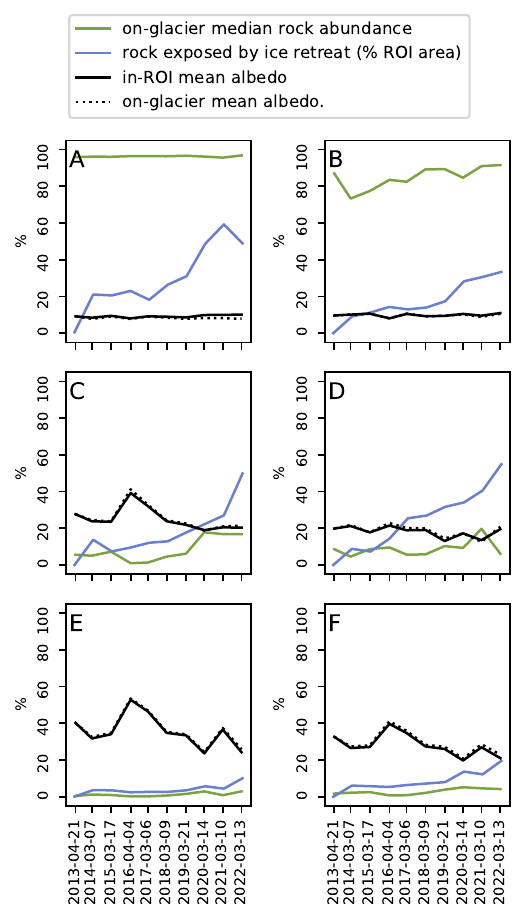}

\caption{Results of multitemporal sediment coverage analysis. Timeseries of four parameters indicated in the legend at the top in the manually chosen ROI:
A -- terminal part of the tongue; 
B -- ablation area below icefalls; 
C -- southern small cirque; 
D -- southeastern section of the western accumulation zone; 
E -- western accumulation zone above 3500 m.a.s.l. in 2013; 
F -- the entire glacier's area. \label{fig:lineGraphs}} 
\end{figure} 

The median on-glacier rock abundance displays varying trends between the ROIs, albeit on-glacier sediment cover and the share of rock exposed by glacial retreat have increased across all ROIs and over the entire glacier since 2013 (Fig.~\ref{fig:lineGraphs}). The in-ROI and on-glacier mean albedo signals are similar to each other in all ROIs. The lower glacier tongue has a low albedo with only a slight difference between in-ROI and on-glacier values (Fig.~\ref{fig:lineGraphs}A,B). The remaining ROIs display a gradual albedo decrease over time, with a punctuated increase in 2016 (Fig.~\ref{fig:lineGraphs}C-F), which is less pronounced in the receding section of the western ablation zone (Fig.~\ref{fig:lineGraphs}D).

The ROIs covering the lowest part of the glacier present the highest median rock abundances, i.e. close to 100\% in the smaller ROI, and an increasing trend above 80\% in the larger one (Fig.~ \ref{fig:resultsMap}A,~B). Compared to the glacier terminus, the sub-icefall zone with areas of exposed ice has a lower proportion of on-glacier sediment (Fig.~\ref{fig:resultsMap}B). The median rock abundance remains high, which can be attributed to both the presence of sediment on the glacier surface or the shrinking of its boundary. 

The western ablation zone and the entire glacier ROIs show a slow, but steady increase in median rock abundance (Fig.~\ref{fig:lineGraphs}E,~F).
The recession and thick debris cover are confined to a small area relative to the entire glacier. Despite rapid changes in these areas, they do not significantly impact the whole-glacier average, as the bulk of the glacier area is composed of exposed ice and snow. Similarly, in the western accumulation zone the observed increase in rock abundance represents only a thin fringe on the margins of the sector, so that the trend is overshadowed by the large number of pixels with no, or near-zero rock abundances (Fig.~\ref{fig:lineGraphs}E).

\subsection{Relationship between rock abundance and albedo and meteorological variables}
 
\begin{figure}[h]
\includegraphics[width=\linewidth]{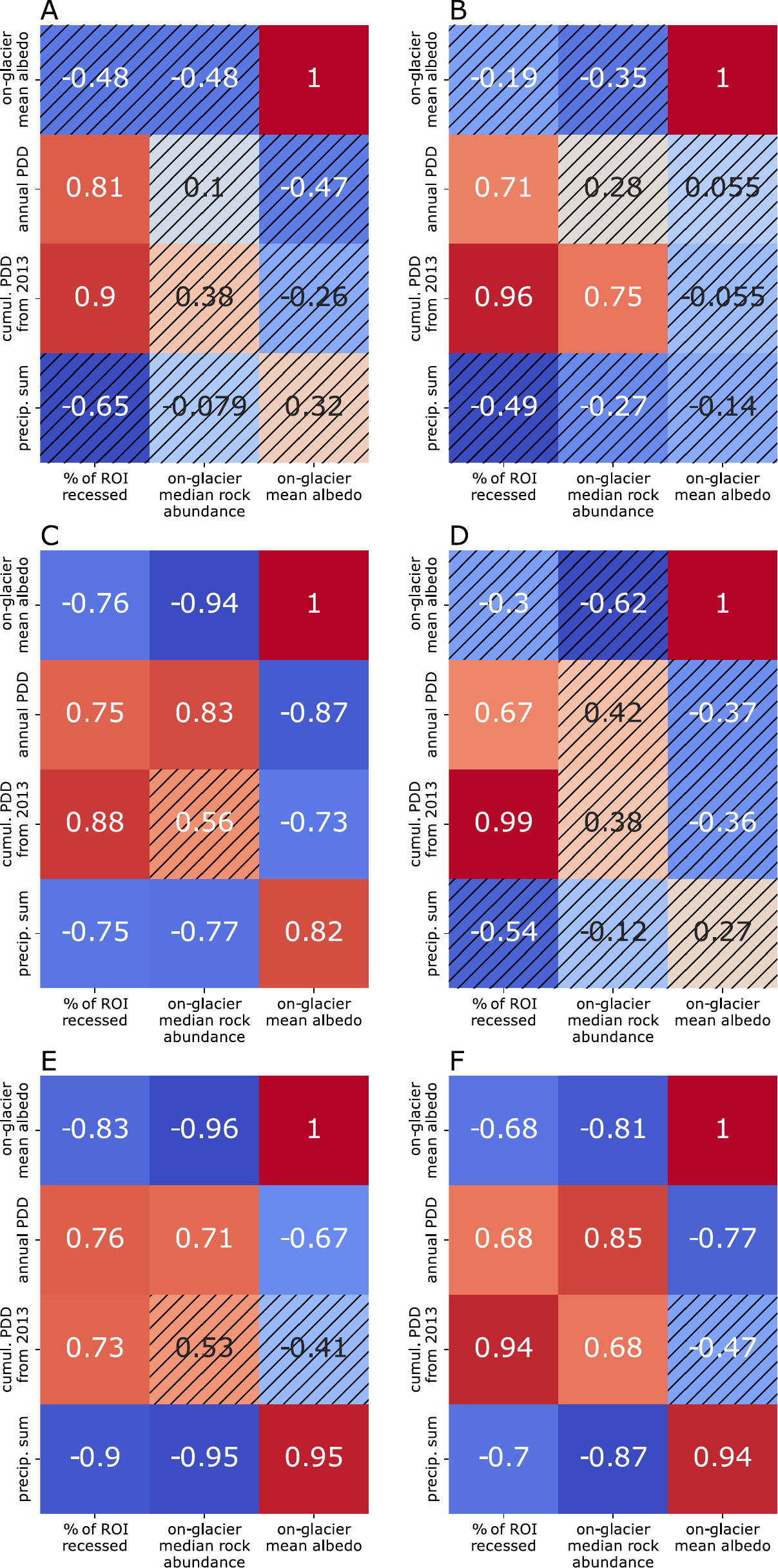}
\caption{Correlation matrices between environmental variables and measures of rock abundance and recession in the selected ROIs. Hatched fields indicate statistically non-significant results.
A -- terminal part of the tongue; 
B -- ablation area below icefalls;
C -- southern small cirque; 
D -- southeastern section of the western accumulation zone; 
E -- western accumulation zone above 3500 m.a.s.l. in 2013; 
F -- the entire glacier's area. 
\label{fig:corrMatrices}}
\end{figure}

Glacier retreat and sediment accumulation, represented by on-glacier rock abundance, are consistently negatively correlated with albedo, in line with the expected darkening effect over areas of retreating ice and on-glacier areas of expanding debris. These correlations are statistically significant in ROIs that do not have intensely retreating margins (Fig.~\ref{fig:corrMatrices}C,~E,~F). The on-glacier rock abundance is the most consistently correlated with albedo.
On the glacier tongue and on the receding part of the eastern accumulation zone the on-glacier albedo is not statistically significantly correlated with rock detection measures (Fig.~\ref{fig:corrMatrices}A,~B,~D). The correlations between on-glacier albedo and rock detection measures are also lower there. 

The only statistically significant correlations in the receding part of the western accumulation zone are between the PDD (both annual and cumulative) and the glacier recession measure (Fig.~\ref{fig:corrMatrices}D). These correlation values are positive and high ($0.67$ and $0.99$). The on-glacier rock abundance does not correlate significantly with any of the environmental measures indicating that in this sector the changes in detected rock abundance are caused by ice margin retreat, not accumulation of debris.

The magnitude of the correlation values are high and statistically significant overall when calculated over the entire 2013 glacier extent (Fig~\ref{fig:corrMatrices}F). Annual and cumulative PDD are both  positively correlated with rock abundance measures, while the mean on-glacier albedo and precipitation sum are negatively correlated with rock abundance. These correlation patterns are shared with the ROIs of the small cirque and the western accumulation (Fig~\ref{fig:corrMatrices}C,~E). The exception is that in the latter, the cumulative PDD does not correlate significantly with the measures other than the \% of exposed bedrock. At the same time, the annual precipitation does correlate highly and significantly with on-glacier median rock abundances signifying the importance of snowfall variability for the accumulation zone (Fig~\ref{fig:corrMatrices}E). 

\subsection{Drainage network} 
\begin{figure}[h]
\centering
\includegraphics[width=.92\columnwidth]{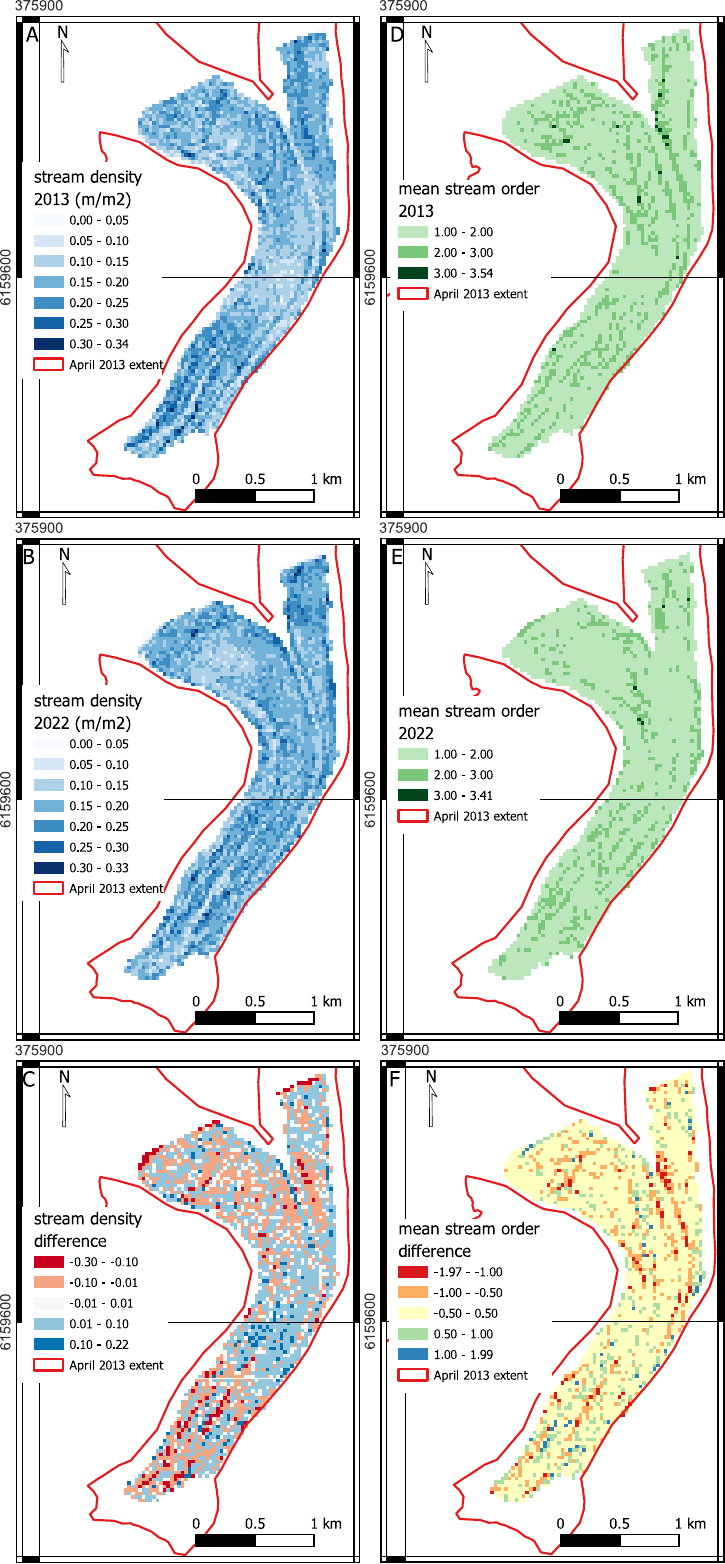}
\caption{Maps of stream network properties and their changes. Left-hand column:  potential stream density as modeled from A -- 2013 LiDAR DEM; B -- 2022 UAV DEM; C -- difference between 2022 and 2013. Right-hand column: mean potential stream order as modeled from D -- 2013 LiDAR DEM, E -- 2022 UAV DEM; F -- difference between 2022 and 2013. \label{fig:streamMaps}}
\end{figure}

Despite the clear evolution towards increasing sediment cover on the glacier (Fig.~\ref{fig:L8debris}), the properties of the glacier tongue drainage network appear to have remained unchanged between 2013 and 2022 (Fig.~\ref{fig:streamMaps}). The maps of potential stream density for both years are visually similar to each other. The observed stream density across the area varies mostly between 0.05 to 0.25~m/m\textsuperscript{2}, reaching values as high as 0.33~m/m\textsuperscript{2} in individual pixels. The stream density map for 2013 has faint outlines of ogives visible in the north-eastern part of the covered sector as alternating bands of higher and lower stream density. In 2022 faint ogive patterns appear lower down the tongue (Fig.~\ref{fig:streamMaps}A-B). In both 2013 and 2022, noticeable sections of medial moraines can be observed at the end of the glacier tongue and along the junction where the two branches of the glacier meet. These sections appear as strips with lower potential stream density compared to the surrounding areas. In 2022, the strip associated with the medial moraine extends further up-glacier compared to 2013.

Changes in potential stream density between 2013 and 2022 range between -0.1 and 0.1~m/m\textsuperscript{2} on most of glacier tongue, with negative values prevalent in the western part of the lower tongue and below the western icefall. Reduced stream density are found over the eastern part of the lower tongue and below the confluence of the two trunks (Fig.~\ref{fig:streamMaps}C). Isolated spots of increasing  stream density of 0.10-0.22~m/m\textsuperscript{2} can also be observed in these areas and along a short, narrow strip along the confluence. Decreases in potential stream density greater than 0.1~m/m\textsuperscript{2} (dark red on Fig.~\ref{fig:streamMaps}C) are also relatively rare but are more spatially clustered than the pixels with extreme increase values (dark blue on Fig.~\ref{fig:streamMaps}C). The most significant decrease in stream density is observed east of the confluence and close to the glacier terminus, particularly along the medial moraine. However, the decrease in stream density does not exceed -0.30m/m\textsuperscript{2} anywhere. The mean stream density change averaged over the entire map is 0.005~m/m\textsuperscript{2}, which is 3.2\% of the 2022 average stream density across the map (0.172~m/m\textsuperscript{2}). In case of stream order, the 2022 average stream order is 1.68 while the change averaged over the whole map is -0.046, a decrease of 2.72\%.

\begin{figure}[htp]
\centering
\includegraphics[width=.92\columnwidth]{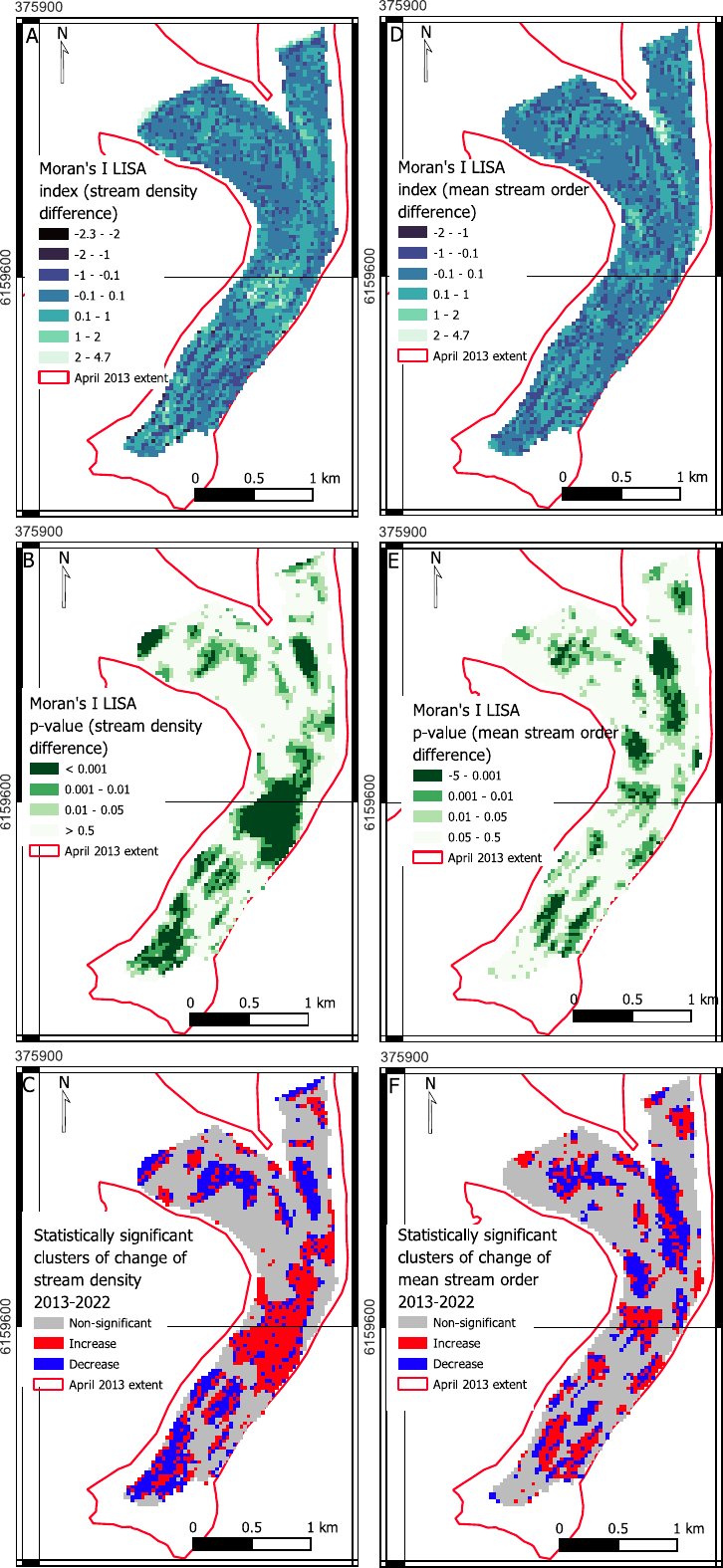}
\caption{Results of Moran I autocorrelation analysis of the stream properties maps for stream density differences (Left-hand column: A, B, C), and stream order differences (right-hand column: D, E, F). The upper row (A, D) shows maps of the local indicators of spatial association (LISA), the middle row (B, E) shows maps with areas of statistical significance of LISA. The lower row (C, F) shows the interpretation of the data in relation to the difference maps.
\label{fig:streamMoran}}
\end{figure} 

Maps of mean stream order show a similar pattern of limited changes during the years between 2013 and 2022 (Fig.~\ref{fig:streamMaps}C-E). The mean stream order across the tongue mostly falls between 1 and 2. Higher values can be observed along the banks of the medial moraines and ogives in 2013, as well as in scattered lines across the lower tongue in 2022. Notably, the cluster of high mean stream order on the western bank of the northern tributary disappeared, which is evident both by comparing the two annual maps (Fig.~\ref{fig:streamMaps}D-E) as well as their differences (Fig.~\ref{fig:streamMaps}F). The difference map demonstrates that stream order were overall stables over time, with few pixels showing an increase above 0.50, while decreases are visible along the medial moraines and the eastern edge of the tongue.

The spatial autocorrelation analysis reveals statistically significant spatial clusters with similar changes in supraglacial hydrology. A region of increasing stream density occurs in the middle of the tongue and extends upwards towards the eastern tributary. Conversely, areas of minimal change in stream density coincide with highly crevassed regions below the western icefall and with areas in the eastern lobe of the lower tongue. Within all of these clusters, there are threads of opposite behavior. 
Notably, spots of stream density increase are observed on sections of ice directly adjacent to the moraines, while in the cluster of densification, which generally exhibits limited debris cover, areas of stream density drops align with the presence of moraines (Fig.~\ref{fig:streamMoran}C). On the map of stream order spatial correlation, the statistically significant clusters indicate an increase in mean stream order in areas of exposed ice and a decrease over moraines (Fig.~\ref{fig:streamMoran}F). 
The stream detection algorithm required to compute maps of sinks (wells) derived from both DEMs. The number of detected wells was similar for both years, with 696 wells detected in 2013 and 690 in 2022. This count includes sinks only within the 2022 glacier contour and excludes the heavily crevassed area in the highest part of the analyzed area. The highly irregular topography in this area leads to the detection of numerous local depressions, but they do not have a significant impact on the stream network outside of their immediate vicinity.

%%%%%%%%%%%%%%%%%%%%%%%%%%%%%%%%%%%%%%%%%%
\section{Discussion}

\subsection{Mapping of glacier facies and ELA estimates}
On the OBIA-classified MS images areas where the snow cover on the map does not extend below the 2013 ELA of 3500~m.a.s.l. are typically located along the mountain slopes, where deep shadows are present in the analyzed images (Fig.~\ref{fig:MSfacies}). However, since the neighboring areas surrounding the shadows are also classified as bare ice, the areas classified as snow-free above 3500~m.a.s.l. cannot be attributed solely to a classification artifact caused by low reflectance in the shaded regions. During the manual mapping of late season snowline extent, examination of the Dove/RE-1 imagery showed patterns of snow cover reappearing in the same locations in years of more abundant snow cover, and with bare ice in the same places in years of limited snowfall. This fluctuation suggests that the white, smooth areas on the imagery were indeed snow overlying ice. These snowy areas were sometimes bright white, but also often appeared darker, indicating a thin layer of dust on top of the snow. 

Albedo measurements averaged across entire glaciers are necessarily biased towards the albedo of the dominant surface type across this glacier~\citep{Naegeli2019}, which for the case of Universidad Glacier would be the bare ice surface. Changes such as ice margin retreat and debris cover expansion on Universidad Glacier occurred on a relatively small portion of the glacier. The percentage of rock exposed within the entire 2013 glacier outline (the whole glacier ROI) increased at a faster rate than within the ROI of the western accumulation area (Fig.~\ref{fig:lineGraphs}E,~F). As the rock abundance remained stable over the years in the accumulation zone ROI, we assume that the change in the albedo measures was largely due to the disappearance of bright snow cover and the exposure of the darker ice beneath.

During the studied period, the AAR dropped from 0.68 to 0.10. Hence more than 80\% of the 2013 accumulation area, corresponding to half of the entire glacier surface, has transitioned from an accumulation regime to an ablation regime within 10 years. During the latter part of this period, between 2019 and 2023, the ELA has shifted up more rapidly to the 3700-4100~m.a.s.l. range, while the AAR shifted down, between 0.48 and 0.04 (Fig.~\ref{fig:ELAgraph}A). When looking at the western accumulation zone the interannual oscillations in ELA and AAR are also reflected in the on-glacier albedo (Fig.~\ref{fig:lineGraphs}E), although the on-glacier rock abundance remained relatively low and stable. However, the strong correlation between on-glacier albedo and on-glacier rock abundance (dust cover, Fig.~\ref{fig:corrMatrices}E)
suggests a synchronized trend between the two, possibly driven by precipitation.

In years of high precipitation, the snow covers the darker bare ice and sediment, while also helping to wash away accumulated dust with meltwater, thereby increasing the glacier albedo. In drier years, increased exposure of glacial ice and unhindered accumulation of dust contribute to a reduction in albedo as these two precipitation-dependent factors work together to darken the surface. 
This conceptual model aligns with a previous study on Chilean glaciers, which identified precipitation and snow cover as the main controlling factors of glacier albedo~\citep{Shaw2020}. The yearly precipitation sum fluctuates alongside the on-glacier albedo, ELA and rock abundance, while the annual PDD sum remained relatively stable from 2020 to 2022 (Fig.~\ref{fig:overviewMap}C). 
The climate-driven fluctuations in AAR and albedo, as well as the low significance of the correlation between rock abundance and cumulative PDD in the western accumulation zone, contribute to the image of the darkening in the accumulation zone being primarily caused by retreat of snow cover, than by accumulation of a layer of debris. 

The visual observations alongside mapping of ELA make it clear that it is glacial ice which is exposed at the end of the ablation season, not a firn layer underlying the snow. The changes in albedo -- and ELA -- we present include disappearance of firn layers in areas where the snow cover has disappeared.

\subsection{Use of GRAI for glacier mapping}
The values of GRAI over Universidad Glacier were consistently negative over both rock and ice. Lower GRAI values correspond to a lower rock abundance (sediment presence) in a pixel, although nearly all pixels show negative GRAI regardless of surface type. This can be compared to a similar index, KBRI, designed to detect karst rock. 
KBRI increases linearly with the increase of exposed bedrock, as described by its proponents~\citep{Pei2018}, with KBRI values above zero observed in pixels with more than 23.9\% rock abundance. Nevertheless, the GRAI values we obtained over exposed rock and moraine do not exceed 0 significantly, and GRAI values of 0 are found for near 100\% of rock abundance (Fig.~\ref{fig:logisticmodel}). This discrepancy can be attributed to different composition of the imaged rocks. The moraines of Universidad Glacier consist of granodiorite~\citep{Lliboutry1958, Fernandez2022}, a type of felsic igneous rock.  Conversely, the original study by~\citet{Pei2018} analyzed karst (carbonate) rocks of Southern China~\citep{Pei2018}. The karst rocks are characterized by a lower reflectance in NIR than in SWIR~\citep{Pei2018}. The granodiorite spectral reflectance curve also exhibits NIR reflectance below SWIR reflectance~\citep{CardosoFernandes2021, Ghrefat2021}. However, the difference between the two bands is lower compared to ice and less pronounced to what was reported for karst by~\citet{Pei2018}. 

The GRAI index is sensitive to the presence of rock material on the surface, allowing for the detection of both sediment cover overlaying the glacier surface and rocky areas exposed due to glacier recession. However, it does not distinguish between these two surface types. Therefore, GRAI can be valuable in determining the extent of ice- and snow-covered sections of glaciers, particularly in relation to efforts focused on tracking recession. This makes it comparable to other index-based glacier mapping approaches~\citep{Haireti2016, Bhardwaj2014, Fleischer2021}. Our comparison between GRAI and albedo highlights the advantage of GRAI in its ability to disregard variations in ice color. Through machine-learning-based classification of high-resolution multispectral imagery, we consistently observed the presence of exposed ice areas on the lower tongue of the glacier. However, these areas appeared very dark due to the presence of embedded sediments~(Fig.~\ref{fig:MSfacies}). The GRAI maps, segmented using unsupervised machine learning, effectively capture this pattern, particularly in the eastern section of the tongue near the medial moraine~(Fig.~\ref{fig:L8debris}). The low correlation between albedo maps and rock abundance in the terminal section of the glacier (Fig.~\ref{fig:corrMatrices}A) supports this finding. The spectrally dark surface of sediment-contaminated ice reflects limited broadband radiation, but it appears distinct from the actual rock surface on a contrast-enhanced infrared index image.

\subsection{Sediment accumulation and glacier retreat}
The increasing rock abundance seen in the Landsat GRAI timeseries (Fig.~\ref{fig:L8debris}C) results from a combination of glacier recession and sediment cover accumulation on the glacier. Different areas of the glacier are subject to ice margin retreat, sediment build-up or both of these processes at once. The gradual exposure of the underlying rock by ice retreat, alongside sediment accumulation, both lead to reduced reflection of incident radiation. In case of the revealed bedrock, this process leads to significant warming of formerly cold areas and the subsequent transfer of the energy to the atmosphere through longwave radiation emitted by the rocks. Consequently, this leads to increased energy flux into the marginal parts of the glacier. This forms a localized feedback loop as increasingly exposed bedrock emits an increasing amount of energy which contributes to acceleration of melting and retreat. When coupled with increasing debris cover thickness this creates two possible scenarios of tongue evolution for glaciers in a setting similar to Universidad's: 1) the very thick sediment cover may insulate the underlying ice from the warm air advecting from the revealed rock. In such case ice thinning and increased melt would be seen on sections of exposed ice, while the debris-covered parts would remain relatively unaffected and emerge progressively higher than their surrounding. 2) Alternatively, a thin layer of accumulated dust may decrease ice albedo and contribute to increased melt on the darker parts, compared to the bright ones -- leading to faster thinning of the darker part. Low variability of on-glacier rock abundance and albedo on the glacier's tongue (Fig.~\ref{fig:lineGraphs}A,~B) as well as low and insignificant correlations between on-glacier albedo, rock abundance and the meteorological variables (Fig.~\ref{fig:corrMatrices}A,~B) indicate a possibility of saturation of the area with debris: a situation when the debris cover is thick enough to prevent detection of any changes. At the same time, a band of revealed, dark ice persists over the years on the eastern part of the tongue (Fig.~\ref{fig:L8debris}C and Fig.~\ref{fig:MSfacies}C). The two scenarios might thus coexists in different areas leading to a heterogenous surfaces and differential melting across the tongue.

Another part of the glacier, located in the former accumulation zone, consists of two small ice cirques to the west of the main trunk. These areas experienced more dynamic changes compared to the larger accumulation zone. In 2013, the two areas were connected to the glacier main trunk through outlets, which transported ice downstream. By February 2023, the two small cirques are visually disconnected from the glacier trunk, although this does not necessarily mean a complete cut-off of ice supply from these sources. The continuous band of dark sediments observed may actually be a thick layer of sediment, beneath which the glacier flow processes are still active. However, the risk of complete disconnection is not the only threat to these two ice bodies. Our mapping indicates a reduction in the extent of these ice areas along their entire perimeter, including in the highest sections (Fig.~\ref{fig:resultsMap}B). Ice margin retreat in this area has accelerated since 2019, leading to the complete cutoff from the main trunk in 2022. The year 2021 was the last year when snowpack was observed in those small cirques at the end of the ablation season (Fig.~\ref{fig:resultsMap}A). It is possible that low precipitation in the year leading up to April 2022 (Fig.~\ref{fig:overviewMap}C) and subsequent times reduced surface runoff during the 2022 ablation season and slowed down processes that could naturally wash down fresh debris from the ice surface. The on-glacier rock abundance detected in this ROI correlates more strongly with precipitation and annual PDD than with cumulative PDD, emphasizing the importance of short-term conditions for debris accumulation.

\subsection{Glacial hydrology}

Significant changes in water-bearing facies in the upper regions of a glacier typically impact the hydrological conditions of glacier sections that drain the accumulation areas. However, deep, wide crevasses located between the accumulation zone and the tongue can capture water that flows downglacier from the accumulation zones and thus divide the glacier into several hydrological zones. 
We hypothesize that this phenomenon mitigates the impact of water flowing from the higher regions of the glacier on the hydrological conditions of the glacier tongue. Meltwater from the upper parts of the glacier drains into englacial and subglacial conduits through crevasses and moulins located at the outlets of the large cirques. As a result, the section of the tongue of the glacier situated below the icefalls represents a distinct hydrological zone. With limited inflow from the upper regions of the glacier, rainfall, local snowmelt, and ice melting are the only significant sources of water on the glacier tongue.  

The glaciers of Chile have experienced negative mass balance in the decade covered by the present work and in the preceding one~\cite{Hugonnet2021b}. The thinning of Universidad Glacier is likely stronger over the tongue, than over higher areas, as implied from a comparison of 2000-2013 ice elevations~\citep{Podgorski2019}. Therefore, we assume that the drainage system in the ablation area undergoes annual re-carving, as intensive melting modifies the preexisting topography every summer, coinciding with the formation of channels by the flowing water. 

The changes of the stream network parameters averaged over the entire map are low, similarly, the number of moulins registered is very close in 2013 and 2022.
This similarity implies a comparable water supply to the system if a network of similar global properties is to be created every year. Such a situation requires either that both precipitation and heat input were at similar levels in 2013 and 2022, or that any change in one factor was counterbalanced by an opposite change in the other.
According to the data obtained from the ERA5-Land reanalysis (Fig.~\ref{fig:overviewMap}C), the cumulative precipitation in the 12 months leading up to 1 April 2013 was 1700~mm, while the corresponding period before 1 April 2022 shows a sum of 975~mm. This indicates that the precipitation value in 2022 was only 57\% of the value observed in 2013. Regarding the cumulative PDD sum, the 12-month period leading up to 1 April 2013 recorded a value of 381\textsuperscript{o}C~d, whereas the same metric for the period before 1 April 2022 was  720\textsuperscript{o}C~d. In this case, the PDD sum in 2022 was 189\% times the value observed in 2013. When recalculated based on the mean elevation of the sub-icefall ROI, the 2022 PDD sum constitutes 178\% of the 2013 PDD sum (844\textsuperscript{o}C~d and 475\textsuperscript{o}C~d, respectively). These numbers highlight the contrasting behavior of the two meteorological factors. The cooler year of 2013 experienced substantial precipitation, while the warmer year of 2022 was characterized by dry conditions. 

Despite the overall similarity, locally, statistically significant changes in stream density and stream order were observed. The measures locally increased over areas of exposed ice, while decreasing over moraines. The decrease in mean stream order and stream density over moraines may result from reduced availability of water from precipitation. Additionally, the debris covered regions could have become more elevated due to differential melting, inhibiting influx of water from upslope. Insufficient water supply in the debris-covered areas inhibits the development of a dendritic network, resulting in drainage confined to singular streams. Meanwhile, the areas of exposed ice experience increases in stream density and mean order, implicating amplified water supply despite reduced precipitation. For this to happen the enhanced ice melt due to rising temperatures would exceed the water loss caused by reduced precipitation. The topographical factor would come into play here as well, as the water from upslope would be channeled onto the sections of bare ice, which are lower than the debris covered ones.

\citet{Bravo2017}~estimated that during the summer 2009/10, Universidad Glacier contributed between 4.3\% and 19.5\% of the water in the Tinguririca river, depending on the month and assumed melt factor. Modeling the melt under different climate change scenarios indicated that the discharge from  Universidad Glacier meltwater stream is projected to reach its maximum around year 2040 and then gradually decrease until 2100~\citep{EscanillaMinchel2020}. The decrease in precipitation due to climatic change is mentioned as the  primary reason for the reduction of water flow in the stream, whereas the peak discharge is attributed to the melting ice from the glacier. \citet{EscanillaMinchel2020} also discovered a clear seasonal distribution of water sources in the stream: precipitation in autumn, snow meltwater in spring, and glacial meltwater during the peak discharge in summer. 

Another study suggests that the ongoing Chilean Mega Drought serves as an analogue for the future conditions, with anthropogenic forcing driving steady decrease in precipitation in the region~\citep{Boisier2018}. In particular, the trend will impact mostly the summer precipitation. Under such conditions, the contribution of glacier to streamflow, including rivers like Tinguririrca, will become even more important for water availability during the summer season. On a broader scale, \citet{Kinnard2020} reported a correlation between the mass balance of Guanaco glacier, located around 600~km north of Universidad Glacier, with the position of the South Pacific Subtropical High. The expansion and migration of this high pressure system were identified as a reason behind the decline in precipitation in this part of South America. The positive correlation between the AAR and the ONI index of the previous winter aligns with the expectation of increased precipitation during El~Ni{\~n}o years~\citep{vanDongen2022}. 

On the other hand, \citet{Barandun2022} highlight factors other than just the climatic forcing  that affect the mass balance of the Paloma Norte and Olivares Alfa glaciers, both located approximately 180~km north of Universidad Glacier. These glaciers suffer from the accumulation of dust on the snow and ice surface due to mobilization of dust from increasingly exposed soils and strip-mining operations. The darkening of the glaciers' surface intensifies melt. While there are no mines in the immediate vicinity, the glacier is surrounded by non-vegetated, rocky slopes. Thus, it can be expected that the decrease in albedo of Universidad Glacier, similarly to the glaciers described by~\citet{Barandun2022}, will contribute to the amplification of melt. 
%%%%%%%%%%%%%%%%%%%%%%%%%%%%%%%%%%%%%%%%%%
\section{Conclusions}

We introduce the Glacial Rock Abundance Index (GRAI), a spectral index based on Landsat 8 OLI data, which is capable of tracking the retreat and expansion of sediment cover on the surface of a mountain glacier. By training a logistic model using a hyperspectral dataset, GRAI values can be converted into fractional rock abundance within a pixel of a Landsat~8~OLI image. GRAI shows the capability to distinguish between darkening caused by the presence of rocky material and darkening caused by the exposure of glacial ice, indicating its superiority over albedo-based debris detection. However, the index is not sensitive to differences between rock material accumulated on the surface of the glacier and the rocky surface exposed due to the retreat of the glacier margin. Further work would be needed to integrate GRAI with thermal-band methods for glacier detection, thereby offering a more precise mapping of sediment accumulation on the Andean glaciers.

Following the warming of the climate in the area during the mega drought period, Universidad Glacier underwent local margin retreat, reduction of the accumulation zone, and uneven accumulation of debris. Within the glacier accumulation zone, the upward progression of the ELA exposed the surface of glacial ice, leading to the absence of seasonally exposed firn across the entire surface of the glacier by the year 2023. This exposed ice has a darker appearance compared to snow. Simultaneously, a thin yet noticeable layer of dust has formed in the accumulation area. The deposition of dust occurs on the surface of ice and snow, a process favored during years with low amount of annual precipitation. The glacier albedo has been steadily decreasing since 2013 due to the combined effects of dust accumulation and the exposure of bare ice, with the latter being the dominant contributing factor.

In the minor cirques to the west of the glacier's primary trunk, both debris accumulation and ice margin retreat have been consistent over the years, posing a potential risk of isolation from the main tongue by 2022-23. The persistent debris layer in these cirques contributes to their diminishing size, possibly leading to their complete disappearance in the near future. These small ice bodies appear more vulnerable to the effects of climate shift compared to the glacial cirques situated in the larger valleys. In the case of the entire Universidad Glacier, this phenomenon results in localized variations in glacier margin retreat. However, in a broader context, the dynamic highlights the importance of increased scrutiny of smaller glaciers. 

Between 2013 and 2022, the Universidad Glacier tongue saw local increases in stream mean order and density in areas of exposed ice, alongside decreases in these two parameters on moraines. These trends point to a significant increase in water flow on the ice surface due to ice melt and reduced water inputs and reduced drainage on exposed moraines, while water flow from melting ice increased significantly. This relationship is consistent with meteorological record derived from reanalysis data which shows a drop in precipitation and a warming between 2013 and 2022. These findings demonstrate direct impacts of climate change which are likely to occur on other Andean glaciers as well. 

%%%%%%%%%%%%%%%%%%%%%%%%%%%%%%%%%%%%%%%%%%

\printcredits

%%%%%%%%%%%%%%%%%%%%%%%%%%%%%%%%%%%%%%%%%%

\section*{Conflicts of interest}
The funders had no role in the design of the study; in the collection, analyses, or interpretation of data; in the writing of the manuscript, or in the decision to publish the~results

\section*{Acknowledgements}

This research was funded by the National Science Center (NCN) PRELUDIUM grant number  2020/37/N/ST10/02481. JP acknowledges partial support from the International Environmental Doctoral School associated with the Centre for Polar Studies at the University of Silesia in Katowice. 
MP acknowledges partial support from the Priority Research Area	(Anthropocene) under the Strategic Programme Excellence Initiative at Jagiellonian University and from ANID through FONDECYT grant 1220978.
The hyperspectral imagery and 2013 elevation data 
were acquired within a project from the Fondo de Innovación para la Competitividad del Ministerio de Economía, Fomento y Turismo, Gobierno de Chile (CK, RU). 
AF acknowledges Fondecyt 1201429 and Anillo ACT210080.
Article contains modified information from ERA5-Land model by Copernicus Atmosphere Monitoring Service 2023. Landsat-8 OLI imagery used in this article courtesy of the U.S. Geological Survey, Planet imagery was provided by the Planet’s Education and Research (E\&R) Program \citep{Planet2022}.

%%%%%%%%%%%%%%%%%%%%%%%%%%%%%%%%%%%%%%%%%%

%% Loading bibliography style file
%\bibliographystyle{model1-num-names}
\bibliographystyle{cas-model2-names}
%\bibliographystyle{elsarticle-harv}

% Loading bibliography database
\bibliography{universidadRef.bib}

\end{document}